\documentclass[12pt]{article}
\usepackage{epsfig,latexsym,colordvi}

\usepackage{graphicx}
\newcommand{\One}{1\kern-4.5pt1}

\newcommand{\be}{\begin{equation}}
\newcommand{\ee}{\end{equation}}

\def\lesim{${\lower 2pt\hbox{$\scriptstyle
<$}\atop\raise 4pt\hbox{$\scriptstyle\sim$}}$} 
\def\grsim{${\lower2pt\hbox{$\scriptstyle >$} \atop\raise4pt\hbox 
{$\scriptstyle\sim$}}$} 
\begin{document}
\begin{center}
\begin{flushright}
     SWAT/05/447\\
October 2005
\end{flushright}
\vskip 10mm
{\LARGE
Supercurrent Flow in NJL$_{2+1}$ at High Baryon Density
}
\vskip 0.3 cm
{\bf Simon Hands and Avtar Singh Sehra}
\vskip 0.3 cm
{\em Department of Physics, University of Wales Swansea,\\
Singleton Park, Swansea SA2 8PP, U.K.}
\vskip 0.3 cm
\end{center}

\noindent
{\bf Abstract:} We present results of numerical simulations of the 2+1$d$
Nambu -- Jona-Lasinio model with non-zero
baryon chemical potential $\mu$ and spatially-varying complex diquark source 
strength $j$. By choosing $\arg(j)$ to vary smoothly through $2\pi$
across the spatial extent of the lattice, a baryon number 
current is induced which in the high density phase remains non-vanishing 
as $\vert j\vert\to0$;
we are hence able to extract a quantity characteristic of a superfluid known as
the {\em helicity modulus}. We also study supercurrent flow at non-zero
temperature and estimate the critical temperature at which the 
normal phase is restored, which is consistent with 
the conventional picture for thin-film
superfluids in which the transition is viewed in terms of vortex -- anti-vortex
unbinding.

\section{Introduction}

There are unfortunately rather few quantum field theories amenable to study
using lattice Monte Carlo techniques in the presence of a non-zero chemical
potential $\mu$, or more specifically with $\mu/T\gg1$. Many important theories,
including QCD, cannot be studied because their path integral measure with
$\mu\not=0$ is not
real on analytic continuation to Euclidean metric, making Monte
Carlo importance sampling inoperative. Of those with positive definite
measure, the Nambu Jona-Lasinio (NJL) model with $N_f=2$ quark
flavors \cite{NJL} is one of the most interesting. At $\mu=0$ the theory
exhibits dynamical chiral symmetry breaking, with the generation of a
constituent quark mass scale $\Sigma$ much larger than the bare mass $m$. 
For $\mu>\mu_c\approx\Sigma$, chiral symmetry is restored, and the ground
state is a degenerate fermi system with $\mu=E_F\simeq k_F$ \cite{SPK}. In 
$d+1$ dimensions the
baryon density in this case is
$n_B=4N_f\mu^d\theta(\mu-\mu_c)/(4\pi)^{d\over2}d\Gamma({d\over2})$.

The precise nature of the ground state at high density depends on $d$. For the
realistic case $d=3$, lattice simulations suggest that
condensation of diquark pairs at the Fermi surface takes place leading to
spontaneous breakdown of the U(1)$_B$ baryon number symmetry \cite{HW}. An
energy gap $\Delta>0$ to excite fermionic quasiparticles develops; for
phenomenologically-motivated lattice parameters the simulations predict
$\Delta/\Sigma\simeq0.15$, in good agreement with self-consistent model
calculations 
of the gap in superconducting quark matter~\cite{CSC}. In this case the NJL
model appears to behave as an orthodox BCS superfluid; there is
long-range ordering of the ground state 
signalled by the non-vanishing condensate 
$\langle qq\rangle\not=0$, and a dynamically-generated mass scale $\Delta$.
Since a U(1)$_B$ symmetry has been spontaneously broken, there is a massless
scalar $qq$
bound state in the spectrum, which is associated with long-range
interactions between vortex excitations in the superfluid, and with a
collective propagating mode for $T>0$ known as second sound.

However, both for obvious numerical convenience, and for a more formal reason,
namely the existence of an interacting continuum limit, lattice simulations were
first applied to the NJL model with $\mu\not=0$ in 
2+1 dimensions~\cite{HM,NJL3prl,NJL3}.
In this case the physics appears radically different. Whilst there is
evidence for long-range coherence of diquark correlation functions~\cite{HM}, 
there is no
long-range order, and apparently no gap. Rather, the condensate
vanishes non-analytically with the diquark source strength, $\langle
qq\rangle\propto j^\alpha$, with $0<\alpha(\mu)<1$~\cite{NJL3prl}. The results
were interpreted in terms of a critical phase for all $\mu>\mu_c$, in which the
diquark correlator decays algebraically, 
$\langle qq(0)qq(\vec r)\rangle\propto r^{-\eta}$ \cite{NJL3}. 
Since all simulations 
are performed on finite systems, and therefore necessarily at $T>0$, the absence
of long-range order is consistent with the Coleman-Mermin-Wagner theorem for
2$d$ systems~\cite{CMW}. The situation is analogous to the
low-$T$ phase of the 2$d$ X-Y model, one of whose physical applications is the
description of superfluidity in thin films~\cite{KT}. 

The defining
property of a superfluid is that the flux density of conserved charge, or
supercurrent $\vec J$, is related to the spatial variation of the phase angle
$\theta$ of the U(1)-valued 
order parameter field (in this case $\langle qq\rangle$) via
\be
\vec J=\Upsilon\vec\nabla\theta.
\label{eq:superJ}
\ee
The constant of proportionality $\Upsilon$ is known as the {\em helicity
modulus}. For a textbook non-relativistic superfluid such as $^4$He 
it is given by 
\be
\Upsilon={\hbar\over M}n_s
\ee
where $M$ is the mass of the helium atom and $n_s$ is a parameter
called the {\em superfluid density\/}, which need not coincide with the charge
density of the condensate. For a relativistic system $\Upsilon$ is best
thought of as a phenomenological parameter in its own right, rather like 
$f_\pi^2$ in $(d+1)$-dimensional chiral model~\cite{HasLeut}. 
One way of understanding 
superfluidity in the absence of long-range order in a 2$d$ system 
is to observe 
that the only way to change the quantised 
circulation $\oint\vec J.d\vec l$ around one
direction of a finite torus is to excite a vortex -- anti-vortex pair, 
and transport
one member of the pair around the other direction of the torus before
re-annihiliation. The energy required to do this scales as $\ln L$ where
$L$ is the size of the system~\cite{KT}: hence in the thermodynamic limit circulation
patterns are topologically stable.

This Letter will present further support for this scenario in NJL$_{2+1}$
by extracting $\Upsilon$ via a calculation of the induced baryon number current
$\vec J=\langle\bar\psi\vec\gamma\psi\rangle$ in response to a spatially varying
diquark source. As well as providing direct verification of superfluid behaviour
in a fermionic model, we will also study the behaviour of $\vec J$ as
temperature $T$ is increased, and find the transition to ``normal'' behaviour at
a critical $T_c$ consistent with analytic expectations.

\section{Method}
The lattice NJL model studied is identical to that of \cite{HM,NJL3}:
\be
S_{NJL}=\sum_x\bar\chi_x
M[\Phi]_{xy}\chi_y+j\chi_x^{tr}\tau_2\chi_x
+\bar\jmath\bar\chi_x\tau_2\bar\chi_x^{tr}+{1\over
g^2}\sum_{\tilde x}\mbox{tr}\Phi^\dagger_{\tilde x}\Phi_{\tilde x},
\label{eq:S}
\ee
where $\chi$, $\bar\chi$ are isodoublet staggered lattice fermion fields, 
$\Phi=\sigma\One+i\vec\pi.\vec\tau$ is an auxiliary bosonic field defined on the
dual lattice sites $\tilde x$, and the
matrix $M$ is
\be
M_{xy}^{pq}=\delta^{pq}\!\!\sum_{\nu=0,1,2}{\eta_{\nu x}\over2}
[e^{\mu\delta_{\nu0}}\delta_{y,x+\hat\nu}-
 e^{-\mu\delta_{\nu0}}\delta_{y,x-\hat\nu}] +\delta_{xy}\left\{m\delta^{pq}
+{1\over8}\sum_{<\tilde x,x>}[\sigma_{\tilde x}\delta^{pq}
+i\varepsilon_x\vec\pi_{\tilde x}.\vec\tau^{pq}]\right\}.
\ee
Here $<\tilde x,x>$ denotes the set of 8 dual sites $\tilde x$ surrounding $x$,
$\eta_{\mu x}=(-1)^{x_0+\cdots+x_{\mu-1}}$ is the Kawamoto-Smit phase
required for a Lorentz covariant continuum limit, and
$\varepsilon_x=(-1)^{x_0+x_1+x_2}$. A full description of the symmetries of
(\ref{eq:S}) and the numerical simulation method is given in
\cite{NJL3}. The only novelty in the current study is that the diquark 
source strengths $j$, $\bar\jmath$ are now specified to be spatially varying, or
``twisted'':
\be
j=j_0\exp(i\theta_{\vec x});\;\;\bar\jmath=j_0\exp(-i\theta_{\vec x})
\ee
with $j_0$ a real constant. To ensure homogeneity and 
single-valuedness on an $L_s^2\times L_t$
lattice we demand
\be
\theta={2\pi\over L_s}(n_1x_1+n_2x_2)\;\;\Rightarrow\;\;\vec\nabla\theta=
{2\pi\over L_s}(n_1,n_2).
\label{eq:gradient}
\ee
A constant supercurrent of the form (\ref{eq:superJ}) 
is therefore specified by a 
pair of integers $(n_1,n_2)$.
It remains to define the conserved baryon number current $J_\nu$:
\be
J_{\nu x}={1\over2}\langle e^{\mu\delta_{\nu0}}\bar\chi_x\chi_{x+\hat\nu}
                 +e^{-\mu\delta_{\nu0}}\bar\chi_x\chi_{x-\hat\nu}\rangle.
\label{eq:J}
\ee
The timelike component of (\ref{eq:J}) is none other than the baryon charge
density $n_B$ reported in \cite{HM,NJL3}. Here we shall use the same stochastic
technique to estimate the quantum expectation value of the spacelike
components $\vec J(j,\mu)=(L_s^2L_t)^{-1}\sum_x\vec J_x(j,\mu)$ 
to demonstrate behaviour of the form
(\ref{eq:superJ}). The strategy will be to compute
$\vec J$ for fixed $(n_1,n_2)$ for a range of $j_0$ 
and extrapolate $j_0\to0$. Behaviour consistent with
(\ref{eq:superJ}) in this limit is deemed to be superfluid.

We used the same simulation parameters as \cite{HM,NJL3}, namely
$g^2=2.0a$, $ma=0.01$, which at $\mu=0$ yields a dynamically-generated
constituent mass, which in effect sets the scale,
of $\Sigma a=0.71$. As $\mu$ is raised, there is a sharp transition from 
a chirally broken vacuum with $\langle\bar\chi\chi\rangle\simeq{2\over
g^2}\Sigma$, $n_B\simeq0$ to a chirally restored phase with 
$n_B>0$ at $\mu_ca\simeq0.65$.
Studies of the fermion dispersion relation in the phase $\mu>\mu_c$
are consistent with a
sharp Fermi surface with $k_F$\lesim$\mu$ and vanishing gap $\Delta\simeq0$
\cite{NJL3prl,NJL3}.

\section{Results}

\subsection{$T=0$}

\begin{figure}[htb]
\begin{center}
\epsfig{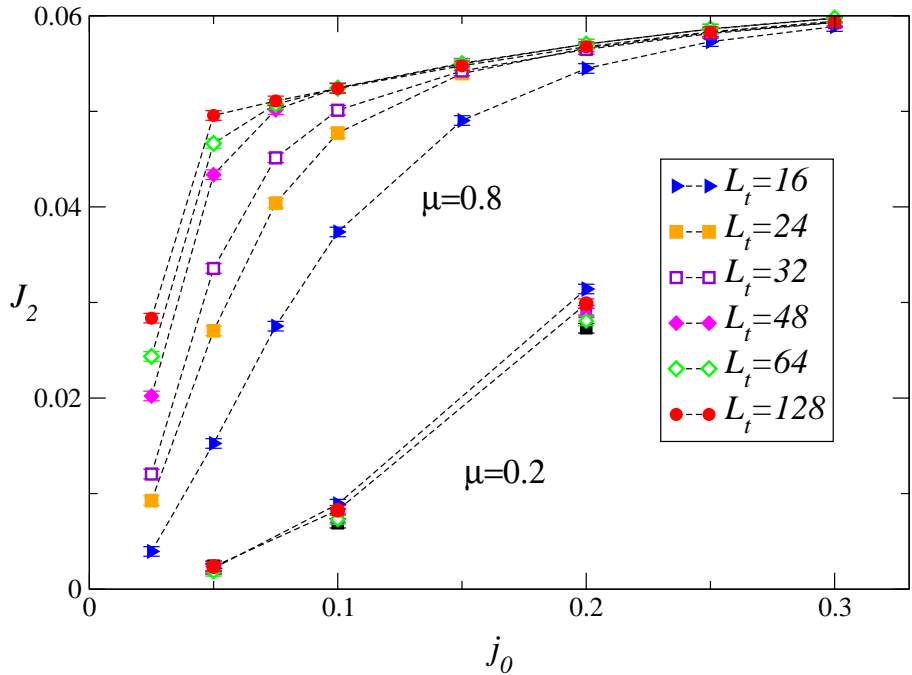}
\end{center}
\vspace{-3mm}
\caption{
$J_2$ vs. $j_0$ on a $16^2\times L_t$ lattice for
two different values of $\mu$.}
\label{fig:Jvsjvart}
\end{figure}
In Fig.~\ref{fig:Jvsjvart} we plot $J_2$ (strictly its imaginary part) 
as a function of $j_0$ for lattices of
various temporal extent $L_t$ at two representative values of $\mu$: $\mu a=0.2$
lies in the chirally-broken low density phase, and $\mu a=0.8$ 
in the high density
phase, where $n_Ba^2\simeq0.25$ \cite{NJL3}.
In all the plots shown here we have chosen $(n_1,n_2)=(0,1)$ to minimise
lattice artifacts, and
from now on we set the lattice spacing $a=1$.

The contrast between the two phases is quite dramatic. For $\mu=0.2$ $J_2$
appears to vary approximately quadratically with $j_0$, and extrapolate to zero
as $j_0\to0$. There is no significant effect as $L_t\to\infty$, or alternatively
as $T\to0$. At $\mu=0.8$ the small-$j_0$ behaviour depends very sensitively 
on $L_t$; as $T\to0$ the data accumulate on a straight 
line which clearly extrapolates to a non-zero value as $j_0\to0$.

This behaviour is readily explained by writing the order
parameter (diquark) field as $\phi=\phi_0e^{i\theta}$, with $\phi_0$ 
approximately constant. A natural
effective Hamiltonian for long wavelength order parameter fluctuations at low
temperature is then
\be
H_{eff}={1\over2}(\vec\nabla\phi)^*.(\vec\nabla\phi)\simeq{\phi_0^2\over2}
(\vec\nabla\theta)^2.
\label{eq:Heff}
\ee
The corresponding Noether current is $\vec
J=-{i\over2}[\phi^*\vec\nabla\phi-(\vec\nabla\phi^*)\phi]
\simeq\phi_0^2\vec\nabla\theta$. For $\mu<\mu_c$,
it is natural to postulate  $\phi$ proportional to $j$, leading
to $J_2(j_0)\propto j_0^2$. For $\mu>\mu_c$, if we
assume that $\lim_{j_0\to0}\phi_0\not=0$ we recover (\ref{eq:superJ}) with
$\Upsilon=\phi_0^2$.

\begin{figure}[htb]
\begin{center}
\epsfig{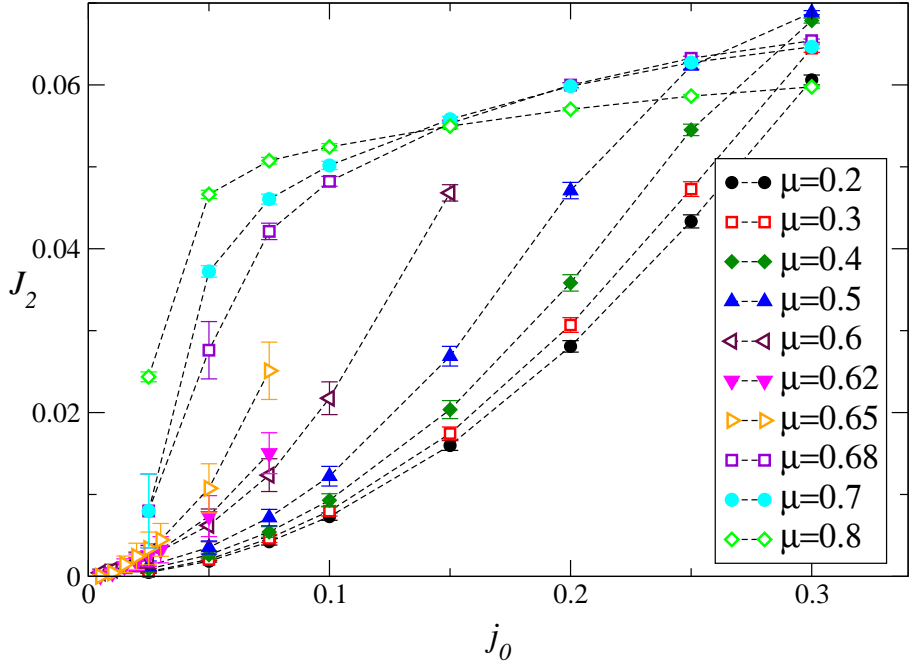}
\end{center}
\vspace{-3mm}
\caption{
$J_2$ vs. $j_0$ on a $16^2\times64$ lattice for
various $\mu$.}
\label{fig:Jvsjvarmu}
\end{figure}
With confidence that $L_t=64$ suffices to determine the phase, 
in Fig.~\ref{fig:Jvsjvarmu} we plot $J_2(j_0)$ for various 
$\mu$. There is a sharp change between values of $\mu\leq0.65$,
which display the low density quasi-quadratic behaviour and smoothly
extrapolate to zero as $j_0\to0$, 
and $\mu\geq0.68$ which show
a negative curvature characteristic of the high density phase. We thus determine
the critical chemical potential for the onset of superfluidity
$0.65<\mu_c<0.68$, in good agreement with the critical value for chiral
symmetry restoration. Since as yet we have no
systematic method of extrapolating to $j_0\to0$ for $\mu$\grsim$\mu_c$
to obtain an estimate for $J_2(\mu)$ as an ``order parameter'', we
can make no decisive statement about the nature of the transition, but note
that the behaviour of $J_2(j_0)$ varies much more sharply across the transition
than the diquark condensate $\langle qq_+(j)\rangle$, either in this model
(Cf. Figure 2 of \cite{NJL3}), or even in NJL$_{3+1}$ (Cf. Figure 4 of
\cite{HW}).  This matches the sharp drop 
in the chiral order parameter
$\langle\bar\chi\chi\rangle$ and corresponding rise in $n_B$ at $\mu=\mu_c$
\cite{HM,NJL3}, and is consistent with the analytic 
prediction of a strong first
order transition in the large-$N_f$ limit \cite{largeN}.

\begin{figure}[htb]
\begin{center}
\epsfig{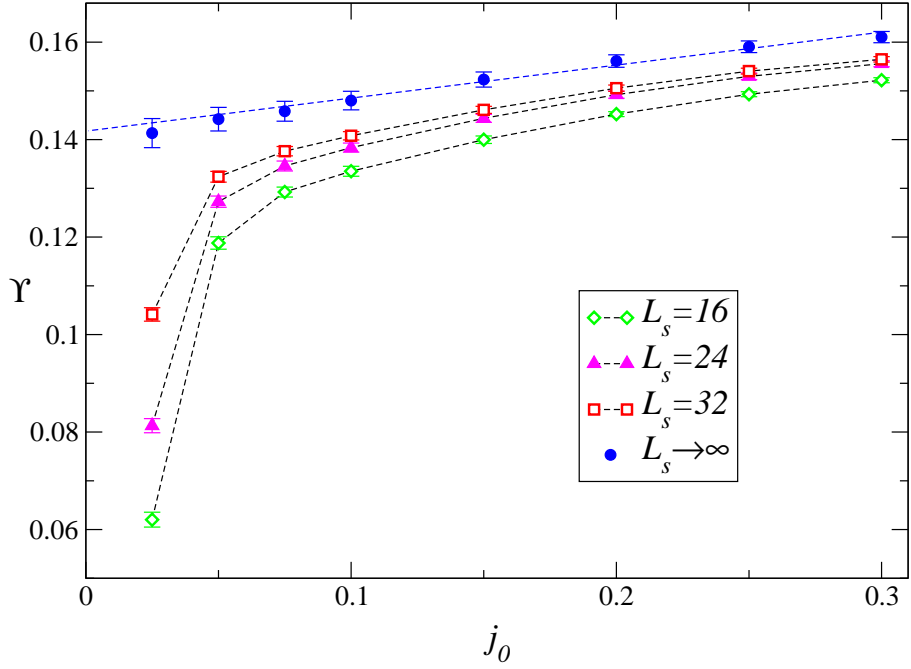}
\end{center}
\vspace{-3mm}
\caption{
$\Upsilon$ vs. $j_0$ on a $L_s^2\times64$ lattice for
$\mu=0.8$.}
\label{fig:Lstoinfty}
\end{figure}
From now on we work exclusively at $\mu=0.8$ in an attempt to understand
the superfluid phase further. In Fig.~\ref{fig:Lstoinfty} we plot
$\Upsilon=J_2L_s/2\pi$ versus $j_0$ on $L_s^2\times64$ lattices. Just as for
the $\langle qq_+(j)\rangle$ data \cite{NJL3}, it turns
out that the data are well-fitted by $\Upsilon(j_0)=A+B/L_s$, resulting in the
$L_s\to\infty$ extrapolation shown in the plot. Recall that as well as
genuine finite-size effects in this case there
may also be some discretisation artifacts, since as $L_s$ increases the 
gradient operator in (\ref{eq:gradient}) becomes better-approximated by the
finite difference. Finally the $\Upsilon(j_0)$ data in the thermodynamic limit 
are extrapolated to $j_0\to0$ with a remarkably
simple linear fit, resulting in
$\Upsilon=0.1413(14)$. We thus quote a result for the helicity modulus 
of $\Upsilon/\Sigma=0.200(2)$ at $\mu
a=0.8$.

It is interesting to pause and ask what value might be expected 
for $\Upsilon$ in a conventional symmetry-breaking scenario. Let
us define diquark operators
$qq_\pm={1\over2}(\chi^{tr}\tau_2\chi\pm\bar\chi\tau_2\bar\chi^{tr})$
and source strengths $j_\pm=j\pm\bar\jmath$, so that the diquark 
terms in the action (\ref{eq:S}) read $j_+qq_++j_-qq_-$.
In the limit $j_-=0$ the equation of
motion for the current is then
\be
\Delta_\mu^- J_\mu=2j_+qq_-.
\label{eq:EoM}
\ee
In the same limit the U(1)$_B$-equivalent of the axial Ward identity
reads
\be
\langle qq_+\rangle=j_+\sum_x\langle qq_-(0)qq_-(x)\rangle
={j_+\over M_-^2}\vert\langle 0\vert qq_-\vert -\rangle\vert^2.
\label{eq:AWI}
\ee
where the second equality 
assumes that the correlation function 
is dominated by a pseudo-Goldstone pole of the form $(k^2+M_-^2)^{-1}$, and
$\vert -\rangle$ denotes a one-Goldstone state. We now introduce the
U(1)$_B$-equivalent form of the PCAC hypothesis:
\be
\langle0\vert\Delta_\mu^-J_\mu\vert-\rangle=\surd\Upsilon M_-^2
=2j_+\langle0\vert qq_-\vert-\rangle
\label{eq:PCAC}
\ee
where we have used the relation $\Upsilon=f_\pi^2$ derived in \cite{HasLeut},
and the second equality follows from (\ref{eq:EoM}). Combining (\ref{eq:AWI})
and (\ref{eq:PCAC}) leads to the equivalent of the ``Gell-Mann-Oakes-Renner''
relation:
\be
\Upsilon_{GMOR} M_-^2=8j\langle qq_+\rangle.
\ee
This can be compared with numerical data for $\langle qq_+(j)\rangle$ and
$M_-(j)$
in \cite{NJL3}. At $j=0.3$, $M_-=0.95$, $\langle qq_+\rangle=0.72$ yielding
$\Upsilon_{GMOR}\approx1.9$; at $j=0.1$, $M_-=0.4$, $\langle qq_+\rangle=0.52$
yielding $\Upsilon_{GMOR}\approx2.6$. We conclude $\Upsilon\ll\Upsilon_{GMOR}$,
consistent with our hypothesis that no symmetry breaking occurs, but
that the dynamics are dominated by long-range phase fluctuations of the order
parameter field, described by a strongly-interacting scalar diquark
field rather than a
weakly-interacting Goldstone mode.

\subsection{$T>0$}

In this section for the first time we explore the superfluid phase at non-zero
temperature. We expect a restoration to the normal phase at some critical
$T_c$. In a comparable numerical study of NJL$_{3+1}$ 
which exhibits superfluidity via
orthodox symmetry breaking \cite{HW}, the value of $T_c$
could be estimated  from the zero temperature gap $\Delta$ via the BCS
prediction
$\Delta/T_c\simeq1.76$. Since this implied that $L_t$ had to exceed $35a$ for 
the system to 
be superfluid, an unambiguous extrapolation $j\to0$ to permit a systematic
study of $T>0$ was not possible. In the current case we shall see that although
the $j\to0$ extrapolation still remains a problem, attaining $T<T_c$ is well
within reach of the simulation.

First let us review a heuristic argument for the expected value of $T_c$,
starting from the Hamiltonian $H_{eff}$ (\ref{eq:Heff}) with
$\phi_0^2=\Upsilon$ \cite{KT}. The phase field $\theta(\vec x)$ may be disrupted
by topologically non-trivial vortex excitations of the form $\theta=q\psi$,
$\vert\vec\nabla\theta\vert=q/r$,
where $q$ is integer and $\vec x$ is written $(r,\psi)$. The energy of a single
vortex is thus
\be
E\approx{\Upsilon\over2}\int_a^{L_s} 2\pi rdr\left({q\over
r}\right)^2=\pi\Upsilon q^2\ln\left({L_s\over a}\right).
\label{eq:E}
\ee
Since a vortex can be located on any of $L_s^2$ lattice sites, the entropy
\be
S=2\ln\left({L_s\over a}\right).
\ee
The free energy $F=E-TS$ thus changes sign for $q=1$ vortices at a critical
temperature 
\be 
T_c={\pi\over2}\Upsilon.
\label{eq:Tc}
\ee
The interpretation 
is that a phase transition separates a low-$T$ superfluid phase
in which vortices are confined to bound dipole pairs, and a
high-$T$ normal phase in which the vortex anti-vortex plasma screens the
long-range interactions responsible for the divergent energy in (\ref{eq:E}).
The relation (\ref{eq:Tc}) remains valid in a more sophisticated
renormalisation group treatment, 
except that $\Upsilon$ is now $T$-dependent and 
should be replaced by its value $\Upsilon(T_c)$ exactly
at the transition \cite{Nelson}.

\begin{figure}[htb]
\begin{center}
\epsfig{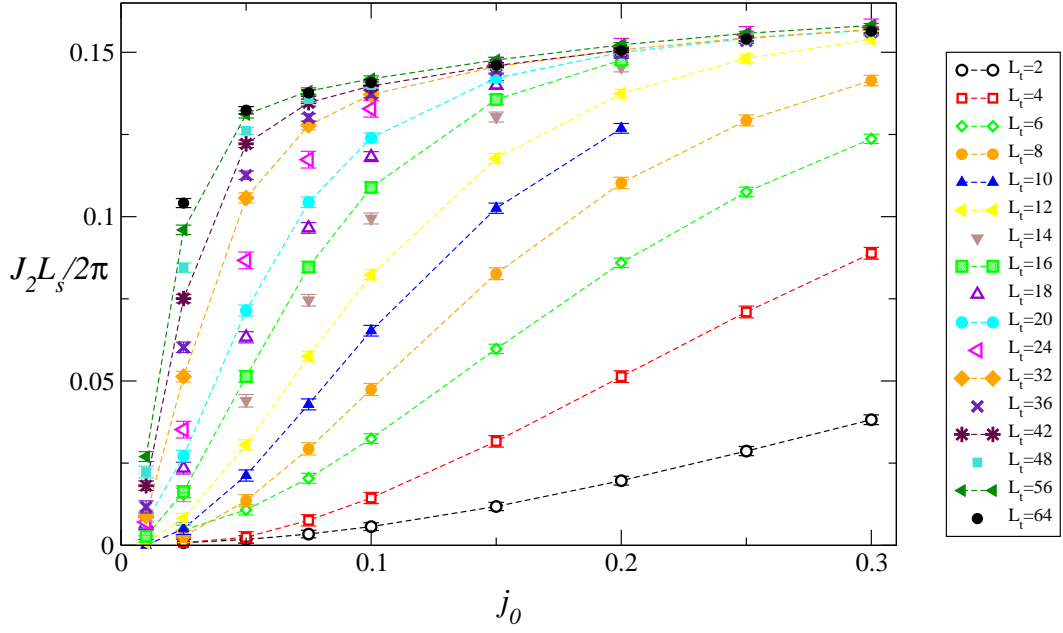}
\end{center}
\vspace{-3mm}
\caption{
$J_2$ vs. $j_0$ on a $32^2\times L_t$ lattice at
$\mu=0.8$ for various $L_t$.}
\label{fig:varyLt}
\end{figure}
Combining our result for $\Upsilon$ with (\ref{eq:Tc}) yields an estimate 
$L_t\approx4.5$ for the temporal lattice extent where a transition to the normal
phase might be expected at $\mu=0.8$. Fig.~\ref{fig:varyLt} shows $J_2(j_0)$ 
on $32^2\times L_t$ lattices with $L_t$ ranging from 64 all the way down
to 2. At the extremes $L_t\geq32$, $L_t\leq4$ the data are reminiscent of 
those characterising respectively the high and low baryon density phases in
Fig.~\ref{fig:Jvsjvart}. 
\begin{figure}[b!h]
\begin{center}
\epsfig{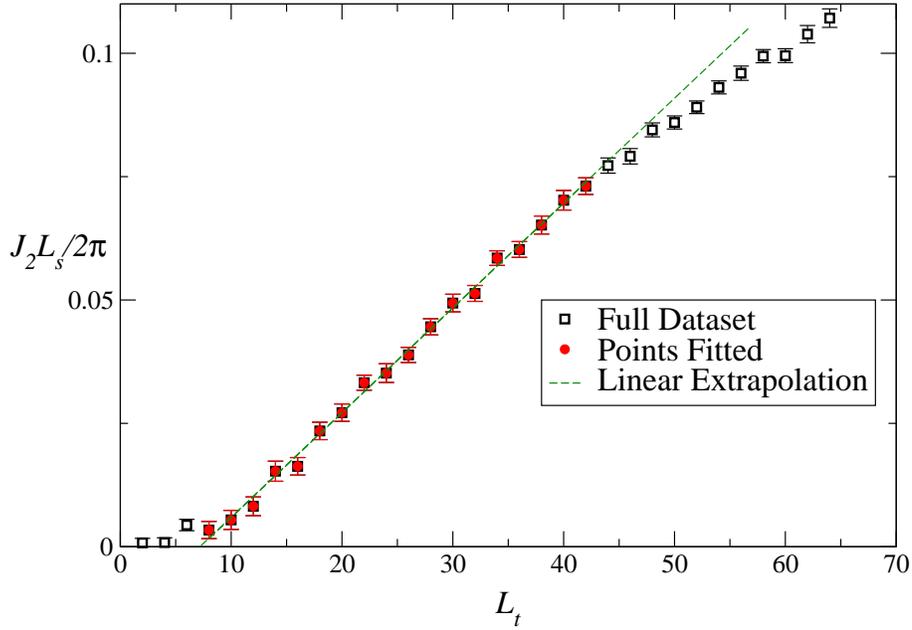}
\end{center}
\vspace{-3mm}
\caption{
$J_2$ vs. $L_t$ on a $32^2\times L_t$ lattice at
$\mu=0.8$ for fixed $j_0=0.025$.}
\label{fig:JvsLt}
\end{figure}
For intermediate temperatures, however, $J_2(j_0)$
shows positive curvature near the origin followed by negative curvature at
larger $j_0$, and once again the means of extrapolating $j_0\to0$ to determine
whether superfluidity persists is unclear.

In Fig.~\ref{fig:JvsLt} we try a different tactic, plotting $J_2$ for every
even $L_t\in[2,64]$ for fixed $j_0=0.025$. 
A linear fit $J_0=a_0L_t+a_1$
through data with $L_t\leq42$ seems quite reasonable, yielding 
$a_0=0.00212(25)$, $a_1=-0.01537(9)$. If we identify the intercept on the
$L_t$-axis with the transition, we deduce $L_{tc}=7.25(95)$ and hence
$\Upsilon/T_c=1.02(13)$, to be compared with the theoretical value 0.637
from (\ref{eq:Tc}). 

\section{Summary}

In this short study of the response of the system to a twisted diquark source
forcing a baryon number current, we have provided direct evidence for the
superfluid nature of the ground state of NJL$_{2+1}$ at high baryon density, 
and quantified it at one representative value of $\mu$  via 
the helicity modulus $\Upsilon$. It should be stressed that the ``physical''
value $\Upsilon/\Sigma\simeq0.2$ quoted is still to be extrapolated to the
continuum limit. It is probably more important to note
that the numerical value of $\Upsilon$ is an order
of magnitude smaller than might be expected in an orthodox symmetry breaking
scenario, and is consistent with the non-Goldstone, strongly self-interacting
nature of the scalar diquark excitations above the ground state. 

We also studied
the response of the system to non-zero temperature. Whilst we were
unable to extrapolate to the zero source limit in a controlled way, by 
studying fixed $j_0$ 
we were able to estimate a critical
temperature $T_c$ for breakdown of superfluidity of the same order of magnitude
as,
and only slightly smaller than, the Kosterlitz-Thouless prediction for a 2$d$
system with U(1) global symmetry, which follows from characterising the 
superfluid/normal transition as arising from vortex pair unbinding.  
More refined
simulations would be required to determine whether $T_c$ actually has the
KT value, or whether NJL$_{2+1}$, which in addition to the scalar diquark
excitations contains 
massless fermion degrees of freedom, 
actually lies in a different universality class, as suggested by
estimates of the critical exponent $\delta$ \cite{NJL3}.

\section*{Acknowledgements}
SJH was supported by a PPARC Senior Research Fellowship. We are grateful to
Costas Strouthos for valuable insight in the early stages of the project.

\end{document}